\documentclass[a4paper]{jpconf}
\usepackage{graphicx}

\usepackage{amssymb}
\usepackage{dcolumn}
\usepackage{amsfonts}
\usepackage{bm}
\begin{document}
\title{Combining nuclear reactions and structure with the dispersive optical model}

\author{W. H. Dickhoff$^1$, M. C. Atkinson$^{1,2}$}
\address{${}^1$ Department of Physics, Washington University in St. Louis, MO 63130 USA}
\address{${}^2$ TRIUMF, BC V6T2A3, CA}


\ead{wimd@physics.wustl.edu,matkinson@triumf.ca}

\begin{abstract}
A review of recent applications of the nonlocal dispersive optical model (DOM) is presented that allows a simultaneous description of nuclear structure and nuclear reactions. An assessment of the quality of the resulting potentials for $^{40}$Ca and $^{48}$Ca is discussed for the description of the $(e,e'p)$ reaction to valence hole states and the possibility of interpreting the data in terms of absolute spectroscopic factors. The relevance of these results in the context of conflicting interpretations between transfer and knockout reactions is pointed out as well as the importance of proton reaction cross sections for isotopes with neutron excess. Application of the nonlocal DOM to $^{48}$Ca incorporates the effect of the 8 additional neutrons and allows for an excellent description of elastic scattering data of both protons and neutrons. The corresponding neutron distribution constrained by all available data generates a prediction for the neutron skin that is larger than most mean-field and available \textit{ab initio} results. Results are presented for the most recent nonlocal DOM analysis of $^{208}$Pb.
\end{abstract}

\section{Introduction}
\label{sec:intro}
How do the properties of protons and neutrons in the nucleus change from the 
valley of stability to the respective drip lines?
The answer can be developed by studying  the propagation of a nucleon through the nucleus at positive energy generating experimentally accessible elastic scattering cross sections as well as the motion of nucleons in the ground state at negative energy.
The latter information sheds light on the density distribution of both protons \emph{and} neutrons relevant for clarifying properties of neutron stars. 
Detailed knowledge of this propagation process allows for an improved 
description of other hadronic reactions, including those that purport to 
extract structure information, like transfer or knockout reactions.
Structure information associated with the removal of nucleons
from the target nucleus, is therefore subject of these 
studies and must be supplemented by the appropriate description of the 
hadronic reaction utilized to extract it.
Consequently, a much tighter link between reaction and structure studies than is common practice is an 
important goal of this research. 
This approach if successful should be able to ultimately resolve the conflicting interpretation of transfer and knockout reactions for rare isotopes~\cite{Gade:2014,Dickhoff:2019}.

Current efforts focus on the Green's functions method~\cite{Dickhoff04,Dickhoff:08} to the nuclear many-body problem to address this issue with special emphasis on reaching the limits of stability. 
The method can be utilized to correlate huge amounts of experimental
data, like elastic nucleon cross sections, analyzing powers, \textit{etc.},
as well as structure information like removal energies, density 
distributions, and other spectral properties.
This is achieved by relating these data to the nucleon self-energy employing its causal properties in the form of a subtracted dispersion relation.
The current implementation and corresponding details can be found in Ref.~\cite{Dickhoff17}. 
The method is known as the dispersive optical model (DOM) and has proceeded way beyond its original form~\cite{Mahaux:1991}.
A more general review of the optical model is available in Ref.~\cite{Dickhoff:2019}.
The problematic status of \textit{ab initio} approaches to the optical model~\cite{Dickhoff:2019} suggests that the DOM approach can act as an interface between \textit{ab initio} theory and experimental data while allowing a sensible and ultimately unambiguous extraction of structure information.
An example of such an analysis is illustrated in Sec.~\ref{sec:eep} where $(e,e'p)$ data are explained utilizing only DOM ingredients which were determined using other experimental data.
We discuss predictions of neutron distributions in Sec.~\ref{sec:skin}, and finally offer some conclusions in Sec.~\ref{sec:con}.

\section{${}^{40,48}$Ca$(e,e'p)^{39,47}$K reactions and spectroscopic factors}
\label{sec:eep}
In the past the relevance of spectroscopic factors has been questioned~\cite{Zhanov10,Furnstahl10}.
It is therefore useful to point out that Fermi liquid theory developed by Landau~\cite{Lan59} relies on the notion of a quasiparticle with a corresponding strength (spectroscopic factor) near the Fermi surface that can be experimentally probed through specific heat measurements~\cite{Wheatley75}.
Experimental efforts in nuclear physics have attempted to extract spectroscopic factors from the $(e,e'p)$ reaction~\cite{Lapikas:1993} for valence hole states in mostly double-closed-shell nuclei (see also Refs.~\cite{Dickhoff:08,Dickhoff:2010}) in suitable kinematic conditions such that a distorted-wave impulse approximation of the cross sections is expected to be valid.

Experimental results of the $(e,e'p)$ reaction have been included in the local DOM in the past by employing the extracted spectroscopic factors~\cite{Kramer:1989,Kramer:2001} in fits with local potentials to the ${}^{40}$Ca and ${}^{48}$Ca nuclei~\cite{Charity:2006,Charity:2007} and to data in other domains of the chart of nuclides~\cite{Mueller:2011}.
A better approach has now been implemented based on the nonlocal DOM developments~\cite{Dickhoff17,Mahzoon:2014,Mahzoon:2017} that also allows an assessment of the quality of the distorted-wave impulse approximation (DWIA) that is utilized to describe the reaction.
We note that the conventional analysis of the reaction employed standard local nondispersive optical potentials to describe the distorted proton waves~\cite{denHerder:1988}.
We have thus arrived at a stage with the DOM that all ingredients for the DWIA description can be supplied from one self-energy that generates the proton distorted waves at the desired outgoing energies, as well as the overlap function with its normalization.
Important to note is that these ingredients are not adjusted in any way to $(e,e'p)$ data.

The nonlocal DOM description of ${}^{40}$Ca data was presented in Ref.~\cite{Dickhoff:2010A,Mahzoon:2014}.
In the mean time, additional experimental higher-energy proton reaction cross sections~\cite{PhysRevC.71.064606} have been incorporated which caused some adjustments of the DOM parameters compared to Ref.~\cite{Mahzoon:2014}. 
Adjusting the parameters from the previous values~\cite{Mahzoon:2014} to describe these additional experimental results leads to an equivalent description for all data except these reaction cross sections.
 Due to the additional absorption at higher energies leads to a loss of strength below the Fermi energy reducing the spectroscopic factors by about 0.05 compared to the results reported in Ref.~\cite{Mahzoon:2014}, thereby also documenting the importance of reaction cross section data for protons at higher energy.

\begin{figure}[tpb]
\begin{center}
\includegraphics[scale=.6]{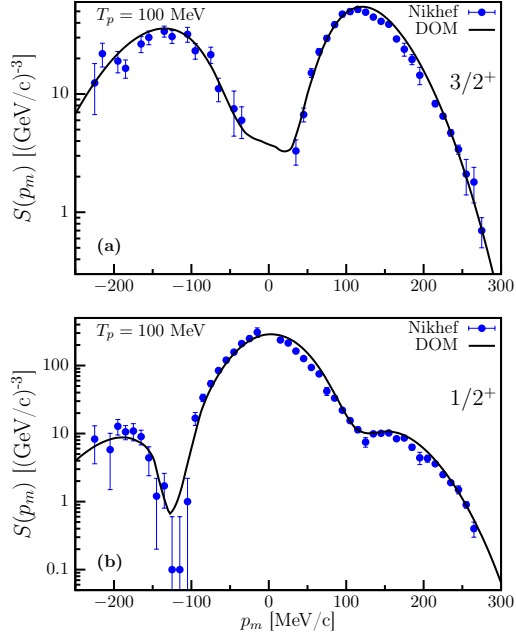}
\caption{Comparison of the spectral distribution measured at Nikhef for outgoing proton energies of 100~MeV to DWIA calculations using the proton distorted waves, overlap function and its normalization from a nonlocal DOM parametrization.
Results are shown for the knockout of a 0$\textrm{d}_{3/2}$ proton from ${}^{40}$Ca to the ground state of ${}^{39}$K.
} 
\label{fig:mack5}
\end{center}
\end{figure}

Using a recent version of the DWEEPY~\cite{Giusti:2011}, 
our DOM ingredients have been utilized to describe the knockout of a proton from the $0\textrm{d}\frac{3}{2}$ and $1\textrm{s}\frac{1}{2}$ orbitals in ${}^{40}$Ca with fixed normalizations of 0.71 and 0.60, respectively~\cite{Atkinson:2018}. 
The DOM at present generates only one main peak for $1\textrm{s}\frac{1}{2}$ orbit so the employed value of 0.60 for the spectroscopic factor takes into account the experimentally observed low-energy fragmentation.
Data were obtained in parallel kinematics for three outgoing proton energies: 100, 70, and 135 MeV.
Data for the latter two energies were never published before.
The resulting description of the $(e,e'p)$ cross sections is at least as good as the Nikhef analysis which yielded spectroscopic factors of 0.65$\pm$0.06 and 0.51$\pm$0.05 for these orbits at 100 MeV~\cite{Kramer:1989} as illustrated in Fig.~\ref{fig:mack5}.
Our results demonstrate that the DWIA reaction model is still satisfactory at 70 MeV and 135 MeV outgoing proton energies~\cite{Atkinson:2018}.
By applying the bootstrap method used for the neutron skin calculation of Ref.~\cite{Mahzoon:2017}, we have generated errors for the spectroscopic factors for these orbits with values 0.71$\pm$0.04 and 0.60$\pm$0.03, for the $0\textrm{d}\frac{3}{2}$ and $1\textrm{s}\frac{1}{2}$ orbitals in ${}^{40}$Ca, respectively.
The results further suggest that the chosen window around 100 MeV proton energy provides the best and cleanest method to employ the DWIA for the analysis of this reaction. 

We therefore make a strong case that the canonical suppression of the spectroscopic factors as pioneered by the Nikhef group~\cite{Lapikas:1993} remains at values of around 0.7 although there are qualitative differences in the build-up of the cross sections on account of the nonlocal potentials that determine the distorted proton waves.
Further insight into the claim that the $(e,e'p)$ reaction can yield absolute
spectroscopic factors for low-lying discrete states in the final 
nucleus~\cite{Sick91,Pandharipande97,Dickhoff:2010} has therefore been provided, while demonstrating that a consistent description of the reaction ingredients as provided by the nonlocal DOM is essential.

The critical question how proton spectroscopic factors depend on adding additional neutrons can be answered by studying the ${}^{48}$Ca$(e,e'p)^{47}$K employing data that were published in Ref.~\cite{Kramer:2001}.
Previously, a fit of $^{48}$Ca was published in Ref.~\cite{Mahzoon:2017}, quoting a neutron skin of $\Delta r_{np} = 0.249\pm0.023$~fm. However, just as in the case of $^{40}$Ca in
Refs.~\cite{Atkinson:2018}, the proton reaction cross section is underestimated at 200 MeV. While there are no experimental data for $^{48}$Ca at these energies, there is a data point at 700 MeV of the
proton reaction cross section of $^{40}$Ca and $^{48}$Ca~\cite{Anderson700}. Comparing the available data for $\sigma_\textrm{react}^{40}(E)$ at 200 MeV and 700 MeV reveals that the reaction cross section essentially stays flat
between these energies. It is reasonable to expect that $\sigma_\textrm{react}^{48}(E)$ assumes the same shape as $\sigma_\textrm{react}^{40}(E)$
at high energies. Thus, data points are extrapolated from the $^{40}$Ca experimental data at energies above 100 MeV by applying the ratio that is seen in the 700 MeV data for
$\sigma_\textrm{react}^{48}(E)/\sigma_{\textrm{react}}^{40}(E)$.
The remainder of the fit did not change significantly from
Ref.~\cite{Mahzoon:2017}. 

To analyze the proton spectroscopic factors, the $^{48}$Ca$(e,e'p)^{47}$K cross section is calculated using the DWIA following the same procedure detailed in
Ref.~\cite{Atkinson:2018} for $^{40}$Ca. 
Just as in Ref.~\cite{Atkinson:2018}, the DOM spectroscopic factors need to be renormalized by incorporating the observed experimental fragmentation of the strength near the Fermi energy that is not yet included in the DOM self-energy. 
This scaling results in a reduction from 0.64 to 0.55 for the $1$s$\frac{1}{2}$ orbital and from 0.60 to 0.58 for the $0$d$\frac{3}{2}$ orbital. These values are in good agreement with
originally published spectroscopic factors~\cite{Kramer:2001} as shown in Table~\ref{table:sf_48}.

\begin{table}[tb]
   \caption[Comparison of spectroscopic factors in $^{48}$Ca]{Comparison of spectroscopic factors in $^{48}$Ca deduced from the previous analysis~\cite{Kramer:2001} using the Schwandt optical potential~\cite{Schwandt:1982} to the normalization of the corresponding overlap functions obtained in the present analysis from the DOM including an error estimate as described in Ref.~\cite{Atkinson:2019}.}
   \vspace{0.5cm}
   \begin{minipage}{\columnwidth}
      \makebox[\columnwidth]{
         \begin{tabular}{ c c c } 
            \hline
            \hline
            $\mathcal{Z}$ & $0\textrm{d}\frac{3}{2}$ & $1\textrm{s}\frac{1}{2}$\\
            \hline
            Ref.~\cite{Kramer:2001} & $0.57 \pm 0.04$ & $0.54 \pm 0.04$\\
            \hline
            DOM & $0.58 \pm 0.03$ & $0.55 \pm 0.03$ \\
            \hline
            \hline
         \end{tabular}
      }
   \end{minipage}
   \label{table:sf_48} 
\end{table}

\begin{figure}[tb]
   \begin{minipage}{\columnwidth}
      \makebox[\columnwidth]{
         \includegraphics[scale=1.0]{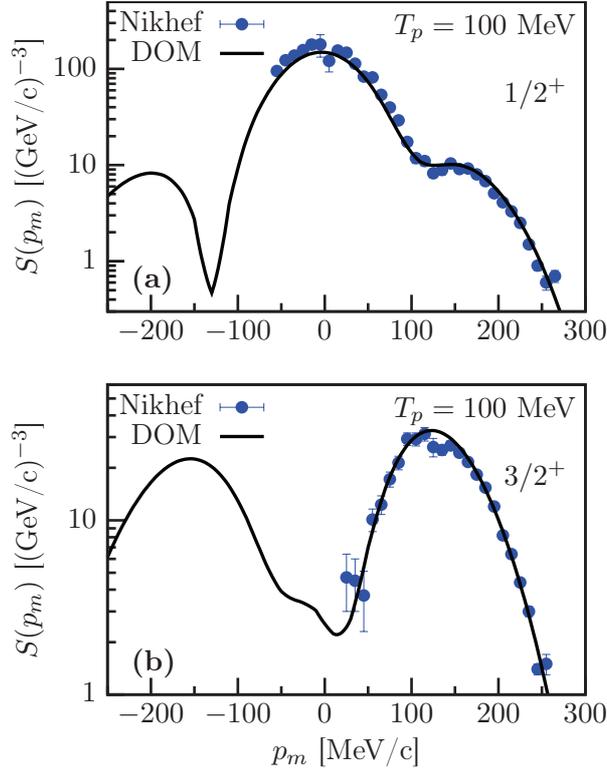}
      }
   \end{minipage}
   \caption[$^{48}$Ca$(e,e'p)$$^{47}$K spectral functions in parallel kinematics at an outgoing proton kinetic energy of 100 MeV]{$^{48}$Ca$(e,e'p)$$^{47}$K spectral functions in parallel kinematics at an outgoing proton kinetic energy of 100 MeV. The solid line is the calculation using the DOM ingredients, while the points are from the 
   experiment detailed in~\cite{Kramer:2001}.
   (a) Distribution for the removal of the $1\textrm{s}\frac{1}{2}$. The curve contains the DWIA for the $1/2^+$ ground state including a DOM generated spectroscopic factor of 0.55 renormalized as described in the text. (b) Distribution for the
   removal of the $0\textrm{d}\frac{3}{2}$ proton with a DOM generated spectroscopic factor of 0.58 for the $3/2^+$ excited state at 0.36 MeV.}
   \label{fig:eep_48}
\end{figure} 
Using the resulting renormalized spectroscopic factors produces the momentum distributions shown in Fig.~\ref{fig:eep_48}.  Thus, the smaller spectroscopic factors in $^{48}$Ca are consistent with the
experimental cross sections of the $^{48}$Ca$(e,e'p)^{47}$K reaction. 
The comparison of $\mathcal{Z}_{48}$ and $\mathcal{Z}_{40}$, the available strength near the Fermi energy, in Table~\ref{table:sf_comp} reveals that both orbitals experience a
reduction. This indicates that strength from the spectroscopic factors is pulled to the continuum at both positive and negative energy, when eight neutrons are added to $^{40}$Ca. 
Thus, the stronger coupling to surface excitations in $^{48}$Ca, demonstrated by the larger proton reaction cross section when compared to $^{40}$Ca, contributes to the quenching of the proton
spectroscopic factors. It is important to note how crucial the extrapolated high-energy proton reaction cross-section data are in drawing
these conclusions. Without them, there is no constraint for the strength of the spectral function at large positive energies, which could result in no quenching of the spectroscopic factors of $^{48}$Ca due
to the sum rule that requires the strength to integrate to 1 when all energies are considered~\cite{Dussan:2014,Dickhoff:08}.

\begin{table}[tb]
   \caption{Comparison of DOM spectroscopic factors in $^{48}$Ca and $^{40}$Ca.
   These factors have not been renormalized and represent the aggregate strength near the Fermi energy.}
   \vspace{0.5cm}
   \begin{minipage}{\columnwidth}
      \makebox[\columnwidth]{
         \begin{tabular}{ c c c } 
            \hline
            \hline
            $\mathcal{Z}$ & $0\textrm{d}\frac{3}{2}$ & $1\textrm{s}\frac{1}{2}$\\
            \hline
            $^{40}$Ca & $0.71 \pm 0.04$ & $0.74 \pm 0.03$ \\
            \hline
            $^{48}$Ca & $0.60 \pm 0.03$ & $0.64 \pm 0.03$ \\
            \hline
            \hline
         \end{tabular}
      }
   \end{minipage}
   \label{table:sf_comp} 
\end{table}

\section{Neutron skin predictions}
\label{sec:skin}

Recently acquired elastic neutron scattering data and total cross sections for ${}^{48}$Ca were published earlier~\cite{Mueller:2011} but it was at that time not possible to generate an accurate fit to the differential cross sections at low energy employing the local implementation of the DOM.
Our current nonlocal DOM potentials provide increased flexibility that allows for the present excellent fit to these data. 
Most of the properties of the first 20 neutrons in this nucleus are already well-constrained by the fit to the properties of ${}^{40}$Ca.
The additional influence of the extra 8 neutrons in this nucleus is then further constrained by these elastic scattering data and total neutron cross sections~\cite{Mueller:2011} as well as level structure.
The neutron properties of ${}^{48}$Ca are of extreme interest to the community since the neutron radius can be experimentally probed without ambiguity employing parity-violating elastic electron scattering experiments at Jefferson Lab~\cite{CREX13}.

\begin{figure}[tpb]
\begin{center}
\includegraphics[scale=.8]{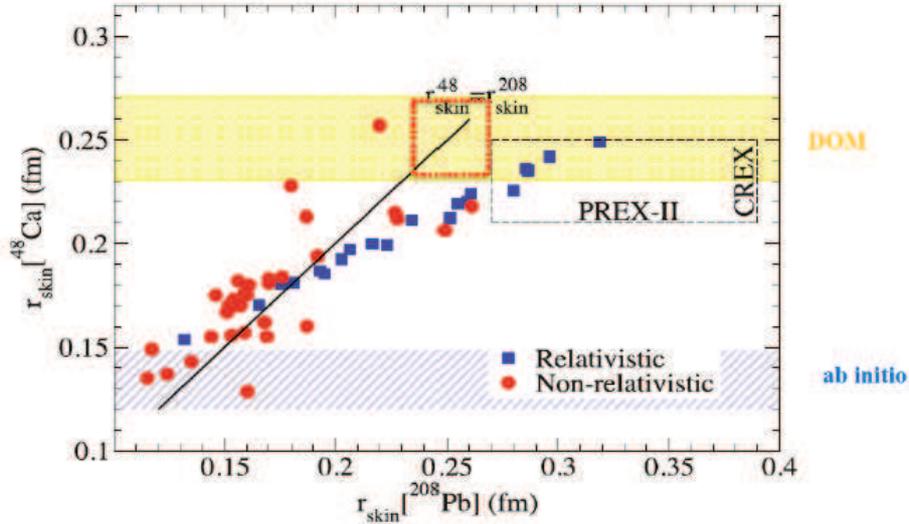}
\caption{Figure adapted from Ref.~\cite{Horowitz14} with the results from Refs.~\cite{Hagen16} and \cite{Mahzoon:2017} indicated by horizontal bars relevant for ${}^{48}$Ca and the big square including the preliminary DOM result for ${}^{208}$Pb~\cite{MackPhD}. Smaller squares and circles refer to relativistic and nonrelativistic mean-field calculations cited in Ref.~\cite{Horowitz14}.}
\label{fig:skin}
\end{center}
\end{figure}

To produce a theoretical error for our result for the neutron skin 
we have employed a method that was explored in the determination of the Chapel-Hill global optical potential~\cite{Varner91}.
For this purpose we employ the experimental errors associated with the neutron elastic scattering data to scramble these results and then generate new fits of these artificial data. 
These results have now been published in Ref.~\cite{Mahzoon:2017} with our neutron skin prediction of 0.249$\pm$0.023 fm which is much larger than the prediction of the \textit{ab initio} coupled-cluster calculation reported in Ref.~\cite{Hagen16} and most mean-field calculations~\cite{Horowitz14}.
We note that this work fulfills the earlier promise of the DOM, in that it can be employed to make sensible predictions of important quantities constrained by other experimental data.
When envisaged earlier~\cite{Charity:2006}, it was thought that these predictions would involve only rare isotopes but important quantities for stable nuclei also fall under its scope.  
We show in Fig.~\ref{fig:skin} results for the neutron skin of ${}^{48}$Ca plotted versus the one of ${}^{208}$Pb including as discussed in Ref.~\cite{Horowitz14} while adding horizontal bars for the DOM result~\cite{Mahzoon:2017} and the coupled-cluster result of Ref.~\cite{Hagen16}.
Our current efforts for ${}^{208}$Pb are also generating a large neutron skin as indicated by the large dotted square in Fig.~\ref{fig:skin}.
The dashed box includes the central value of the Ref.~\cite{PhysRevLett.108.112502} but with the expected error of the new PREX-II experiment. 
The expected error for the CREX experiment~\cite{CREX13} is indicated by the vertical width of the box while its central value is arbitrarily chosen.

\section{Conclusions}
\label{sec:con}
The present nonlocal DOM analysis of the ${}^{40}$Ca$(e,e'p)^{39}$K and ${}^{48}$Ca$(e,e'p)^{47}$K reactions  demonstrates a distinct reduction of the spectroscopic proton strength when 8 neutrons are added to $^{40}$Ca.
These results are in agreement with the trend observed in Ref.~\cite{Gade:2014} but with a reduced slope and in disagreement with the recent analysis reported in Refs.~\cite{Aumann18,Kawase:2018} and with the results of transfer reactions~\cite{Dickhoff:2019}. Some form of quenching is inevitable if one accepts the $np$ dominance picture of short-range correlations~\cite{Rios:2009,Rios:2014,Hen:2017,Duer:2018}, since the added neutrons cause the protons to become more correlated. The increase in the
high-momentum content of protons in $^{48}$Ca is consistent with the $np$ dominance picture, hence it contributes to the quenching of the spectroscopic factors. Additionally, the increased proton reaction
cross section of $^{48}$Ca at all energies compared to $^{40}$Ca leads to more depletion, which also contributes to the observed quenching.  The proton reaction cross section plays a delicate role in determining the
spectroscopic factor.  While in the case of $^{48}$Ca the lack of proton reaction
cross-section data points at energies between $100$-$200$ MeV was compensated for by modifying the corresponding $^{40}$Ca data
points~\cite{Atkinson:2019}, precise measurements of the proton reaction cross sections at these energies are crucial in constraining spectroscopic factors.
Such measurements in inverse kinematics with rare isotopes can further help understand the behavior of spectroscopic factors away from the valley of stability.

Results for the neutron skin in ${}^{48}$Ca and ${}^{208}$Pb indicate that the nonlocal implementation of the DOM generates predictions that are larger than all nonrelativistic and most relativistic mean-field results as well as the \textit{ab initio} result of Ref.~\cite{Hagen16}. 
New PREX and CREX data may further clarify the validity of our interpretation.

\section*{Acknowledgements}
   This work was supported by the U.S. National Science Foundation under grants PHY-1613362 and PHY-1912643.

\section*{References}
\bibliographystyle{iopart-num}
\bibliography{INPC}
\end{document}